\documentclass[prl,twocolumn,floats,showpacs,preprintnumbers,tighten,epsfig,superscriptaddress]{revtex4}
\usepackage{graphicx}
\usepackage{dcolumn}
\usepackage{bm}
\usepackage{amsmath}
\usepackage{amsfonts}
\usepackage{amssymb}

\begin{document}

\title{Spatial pair correlations of atoms in molecular dissociation}
\author{C.~M. Savage}
\affiliation{ARC Centre of Excellence for Quantum-Atom Optics,
Department of Physics, Australian National University, Canberra ACT
0200, Australia}
\author{K.~V. Kheruntsyan}
\affiliation{ARC Centre of Excellence for Quantum-Atom Optics, School of Physical
Sciences, University of Queensland, Brisbane, QLD 4072, Australia}
\date{\today}

\begin{abstract}
We perform first-principles quantum simulations of dissociation of
trapped, spatially inhomogeneous Bose-Einstein condensates of
molecular dimers. Specifically, we study spatial pair correlations
of atoms produced in dissociation after time of flight. We find that
the observable correlations may significantly degrade in systems
with spatial inhomogeneity compared to the predictions of idealized
uniform models. We show how binning of the signal can enhance the
detectable correlations and lead to the violation of the classical
Cauchy-Schwartz inequality and relative number squeezing.
\end{abstract}

\maketitle

Producing and utilizing quantum mechanical correlations between
entangled particle pairs are major themes in quantum mechanics
research. In this paper we investigate correlations of the kind used
by Einstein, Podolsky and Rosen (EPR) to argue that local realism is
inconsistent with quantum mechanics being a complete theory of
nature \cite{EPR}. EPR type non-local correlations were first
produced experimentally in the form of photon pairs from
nondegenerate parametric down-conversion \cite{Ou}. Here, the
correlations are observed between the optical quadrature amplitudes
of spatially separated signal and idler beams, which are analogous
to canonical position and momentum variables originally discussed by
EPR.

It should be possible to extend this type of experiment to
ultra-cold atoms, which would allow the investigation of EPR
correlations for particles with non-zero rest mass. These might be
produced by the transfer of correlations from photons to atoms \cite{Light-to-Atoms}, or else
by nonlinear matter-wave interactions.
The latter include atomic four-wave mixing via Bose-Einstein condensate (BEC) collisions \cite%
{Pu-Meystre-Duan,4WM-theory} and matter-wave down-conversion
via dissociation of a molecular BEC \cite%
{Kurizki-Moelmer2001,twinbeams,Yurovsky,EPR-dissociation}.
 Although the direct measurement of
matter-wave quadrature correlations of this type is a challenging
task  \cite{EPR-dissociation}, a simpler
measurement of density-density correlations -- as a
prerequisite for the EPR quadrature correlations -- can be performed
with current experimental techniques.

In this paper we perform first-principles quantum simulations of
dissociation of a BEC of molecular dimers into pair-correlated
bosonic atoms. This has been demonstrated
experimentally using ultracold (but not Bose condensed)
$^{87}$Rb$_{2}$ dimers \cite{Durr}, although the atom
correlations have not yet been measured. In the case of fermionic atoms \cite%
{Jack-Pu,Fermidiss}, such as in dissociation of $^{40}$K$_{2}$ dimers \cite%
{Greiner}, the density-density correlations between the atoms in two
different spin states have been measured; this case, however,
currently resists first-principles quantum simulations.
Nevertheless, many insights  from the present
bosonic case  also apply to the fermionic case, and to correlations
produced via BEC four-wave mixing \cite{Orsay}. For
example; the importance of operationally well defined measures of
correlations, the role of spatial inhomogeneity, and the trade-off
between image resolution and binning to enhance correlation
detection.

Measurements now challenge theory to give precise
quantitative descriptions of the observable correlations.
Experiments measuring atom shot noise in absorption images \cite%
{Altman-Lukin,Greiner,Bloch} and using
microchannel plate detectors \cite{MPC} have demonstrated atom-atom correlations
\cite{Greiner,Orsay}. However, their quantitative theoretical analysis is lacking.
There is also a need to
determine the best operational measures
\cite{Fermidiss,moldisspaper1} of correlations for these
experiments. The present work  is a step
towards these goals, and a benchmark for other (approximate) theories.

Previous theoretical work on molecular dissociation has explored the atomic
correlations in simplified cases such as 1D systems \cite%
{twinbeams,EPR-dissociation}, without depletion of the molecules
during conversion \cite{Fermidiss}, or for spatially uniform systems
\cite{Jack-Pu,moldisspaper1}. The present work builds on the
characterization of the atomic correlations in momentum space in
Ref.~\cite{moldisspaper1}. Here, we extend the analysis to spatial
correlations and study molecular dissociation in 3D under
experimentally realistic conditions, including the effects of
inhomogeneity of the initial molecular BEC \cite{Vardi} and
time-of-flight expansion. We focus on the use of absorption images,
obtained by line of sight integration, to measure the atomic
correlations.

To investigate the quantum dynamics of molecular dissociation we use
stochastic equations in the positive $P$-representation \cite{+P}.
In this method, one simulates the evolution of four
complex stochastic ($c$-number) fields $\Psi _{i}(\mathbf{x%
},t)$ and $\Phi _{i}(\mathbf{x},t)$ [$\Phi _{i}^{\ast }(\mathbf{x}%
,t)\neq \Psi _{i}(\mathbf{x},t)$], representing the
field operators $\hat{\Psi}_{i}(\mathbf{x},t)$ and $\hat{\Psi}%
_{i}^{\dag }(\mathbf{x},t)$, respectively, where $i=0$ stands for
molecules and $i=1$ for atoms. Averages of the stochastic field
products over a large number of trajectories $\langle \ldots \rangle
_{\mathrm{st}}$ correspond to quantum mechanical ensemble averages
of normally-ordered operator moments. For example $\langle \lbrack \hat{\Psi}%
_{i}^{\dagger }(\mathbf{x},t)]^{k}[\hat{\Psi}_{j}(\mathbf{x}^{\prime
},t)]^{n}\rangle =\langle \lbrack \Phi _{i}(\mathbf{x},t)]^{k}[\Psi _{j}(%
\mathbf{x}^{\prime },t)]^{n}\rangle _{\mathrm{st}}.$

The stochastic differential equations governing the quantum dynamics
of dissociation are \cite{moldisspaper1}
\begin{align}
\frac{\partial \Psi _{1}}{\partial t}& =\frac{i\hbar }{2m_{1}}\mathbf{\nabla
}^{2}\Psi _{1}-i\ \Delta \Psi _{1}+\chi \Psi _{0}\Phi _{1}+\sqrt{\chi \Psi
_{0}}\;\zeta _{1},  \notag \\
\frac{\partial \Phi _{1}}{\partial t}& =-\frac{i\hbar }{2m_{1}}\mathbf{%
\nabla }^{2}\Phi _{1}+i\Delta \Phi _{1}+\chi \Phi _{0}\Psi _{1}+\sqrt{\chi
\Phi _{0}}\;\zeta _{2},  \notag \\
\frac{\partial \Psi _{0}}{\partial t}& =\frac{i\hbar }{2m_{0}}\mathbf{\nabla
}^{2}\Psi _{0}-\frac{\chi }{2}\Psi _{1}^{2},  \notag \\
\frac{\partial \Phi _{0}}{\partial t}& =-\frac{i\hbar }{2m_{0}}\mathbf{%
\nabla }^{2}\Phi _{0}-\frac{\chi }{2}\Phi _{1}^{2}.  \label{Positive-P-eqs}
\end{align}%
Here, $\zeta _{j}(\mathbf{x},t)$ ($j=1,2$) are real,
independent Gaussian noises with zero means and nonzero correlations $\langle \zeta _{j}(%
\mathbf{x},t)\zeta _{k}(\mathbf{x}^{\prime },t^{\prime })\rangle _{\mathrm{st%
}}=\delta _{jk}\delta (\mathbf{x}-\mathbf{x}^{\prime })\delta
(t-t^{\prime })$, $m_{1}$ and $m_{0}=2m_{1}$ are the atomic and
molecular masses, $\Delta $ is the detuning corresponding to the
energy mismatch $2\hbar \Delta =2E_{1}-E_{0}$ between the free
two-atom state at the dissociation threshold and the bound molecular
state, and $\chi $ is the coupling responsible for coherent
conversion of molecules into atom pairs,
e.g., via Raman transitions or a Feshbach resonance (see Refs. \cite%
{moldisspaper1,Raman-Feshbach} for details).

We have ignored $s$-wave scattering interactions because
they do not significantly affect the dynamics or the
correlations for durations resulting in less than $\sim
10\%$ conversion \cite{moldisspaper1}. The characteristic timescale
for this can be estimated using a Fermi's golden rule calculation
\cite{Davis}, giving $t\lesssim 0.1\Gamma ^{-1}$, where $\Gamma
=\chi ^{2}(m_{1}/2\hbar )^{3/2}\sqrt{|\Delta |}/\pi $ is the
molecular decay rate. The duration of dissociation should also be
shorter than the timescale for losses due to inelastic collisions \cite{Rempe-inelastic}.

We assume that the molecular BEC is initially in a coherent state,
whereas the atoms are in the vacuum state. Once the dissociation is
suddenly switched on \cite{Hanna}, the trapping potentials are
simultaneously switched off, so that the evolution takes place in
free space. For molecules at rest, the excess of potential energy is
converted into kinetic energy, $2\hbar |\Delta |\rightarrow 2\hbar
^{2}k^{2}/(2m_{1})$, of dissociated atom pairs with equal but
opposite momenta around $\pm \mathbf{k}_{0}$, where $
|\mathbf{k}_{0}|=\sqrt{2m_{1}|\Delta |/\hbar }$. This is the
physical origin of the expected correlations between the atoms.
Ideally, in the time-of-flight expansion the momentum correlations
are converted into position correlations between diametrically
opposite atoms in the far field. In non-ideal cases, such as in
strongly inhomogeneous systems with large momentum uncertainty, or
with insufficient time for expansion, the spatial correlations may
degrade.

In absorption imaging the number of photons detected by each camera pixel
determines the number of atoms contained in the volume of a narrow column in
the imaging laser's propagation direction, which we denote as $z$.
Denoting the area of the camera pixel about the point $\mathbf{r\equiv x}%
_{\perp }=(x,y)$ on the detection plane as $A(\mathbf{r})$, the
corresponding atom number operator is given by
\begin{equation}
\hat{N}_{\mathbf{r}}=\int_{A(\mathbf{r})}d\mathbf{r}^{\prime }\,\int dz\,%
\hat{n}(\mathbf{r}^{\prime },z)=\int_{A(\mathbf{r})}d\mathbf{r}^{\prime }\,%
\hat{n}_{\perp }(\mathbf{r}^{\prime }) .  \label{pixel_number}
\end{equation}%
On a computational grid, the number operator
$\hat{N}_{\mathbf{r}}$ is
related to the integrated 2D column density $\hat{n}_{\perp }(\mathbf{r}%
)=\int dz\,\hat{n}(\mathbf{x})$ via $\hat{N}_{\mathbf{r}}=\hat{n}_{\perp }(%
\mathbf{r})\Delta x\Delta y$, where $\hat{n}(\mathbf{x})=\hat{\Psi}%
_{1}^{\dagger }(\mathbf{x})\hat{\Psi}_{1}(\mathbf{x})$ is the 3D
density, $\Delta x$ and $\Delta y$ are the lattice spacings, and
$A(\mathbf{r})=\Delta x\Delta y$.

Correlation between the atom number fluctuations in a pair of different
pixels can be quantified via the normalized number-difference variance ($\mathbf{%
r}\neq \mathbf{r}^{\prime }$)
\begin{equation}
V_{\mathbf{r},\mathbf{r}^{\prime }}=\frac{\langle \lbrack \Delta (\hat{N}_{%
\mathbf{r}}-\hat{N}_{\mathbf{r}^{\prime }})]^{2}\rangle }{\langle \hat{N}_{%
\mathbf{r}}\rangle +\langle \hat{N}_{\mathbf{r}^{\prime }}\rangle }=1+\frac{%
\langle :[\Delta (\hat{N}_{\mathbf{r}}-\hat{N}_{\mathbf{r}^{\prime
}})]^{2}:\rangle }{\langle \hat{N}_{\mathbf{r}}\rangle +\langle \hat{N}_{%
\mathbf{r}^{\prime }}\rangle },  \label{variance}
\end{equation}%
where $\Delta \hat{C}=\hat{C}-\langle \hat{C}\rangle $ is the
fluctuation in $\hat{C}$ and the colons $::$ indicate
normally-ordered operator products. This definition uses the
conventional normalization with respect to the shot-noise level
of
Poissonian statistics, such as for a coherent state, $%
\langle \hat{N}_{\mathbf{r}}\rangle +\langle
\hat{N}_{\mathbf{r}^{\prime }}\rangle $.
$V_{\mathbf{r},\mathbf{r}^{\prime }}=1$ for uncorrelated signals.
Variance smaller than one,
$V_{\mathbf{r},\mathbf{r}^{\prime }}<1$, implies reduction (or
squeezing) of fluctuations below the shot-noise level and is
due to correlation between particle number fluctuations in the $%
\mathbf{r}$ and $\mathbf{r}^{\prime }$ pixels. Perfect ($100$\%)
squeezing of the number-difference fluctuations corresponds to
$V_{\mathbf{r},\mathbf{r}^{\prime }}=0$.

The number-difference variance is related to Glauber's second-order
correlation function $g^{(2)}(\mathbf{r},\mathbf{r}^{\prime })=\left\langle :%
\hat{n}_{\perp }(\mathbf{r})\hat{n}_{\perp }(\mathbf{r}^{\prime
}):\right\rangle /(\langle \hat{n}_{\perp }(\mathbf{r})\rangle \langle \hat{n%
}_{\perp }(\mathbf{r}^{\prime })\rangle )$. In the simplest symmetric case,
with $%
\langle \hat{n}_{\perp }(\mathbf{r})\rangle =\langle \hat{n}_{\perp }(%
\mathbf{r}^{\prime })\rangle $ and $g^{(2)}(\mathbf{r},\mathbf{r})=g^{(2)}(%
\mathbf{r}^{\prime },\mathbf{r}^{\prime })$, the relationship is
\begin{equation}
V_{\mathbf{r},\mathbf{r}^{\prime }}=1+\langle \hat{N}_{\mathbf{r}}\rangle
\lbrack g^{(2)}(\mathbf{r},\mathbf{r})-g^{(2)}(\mathbf{r},\mathbf{r}^{\prime
})].  \label{V-g}
\end{equation}%
In order that $V_{\mathbf{r},\mathbf{r}^{\prime }}$ is suppressed
below the level of classical, uncorrelated statistics, $V_{\mathbf{r},%
\mathbf{r}^{\prime }}<1$, it is necessary and sufficient to have $g^{(2)}(%
\mathbf{r},\mathbf{r}^{\prime })>g^{(2)}(\mathbf{r},\mathbf{r})$
which is a direct signature of the violation of the classical
Cauchy-Schwartz inequality \cite{Reid-Walls}. Similarly, $V_{\mathbf{r},%
\mathbf{r}^{\prime }}\geq 1$, implies $g^{(2)}(%
\mathbf{r},\mathbf{r}^{\prime })\leq g^{(2)}(\mathbf{r},\mathbf{r})$
which agrees with the classical Cauchy-Schwartz inequality. Thus,
Eq. (\ref{V-g}) shows that to have a non-trivial
quantum correlation it is not enough to have $g^{(2)}(%
\mathbf{r},\mathbf{r}^{\prime })>1$, but one has to have $g^{(2)}(%
\mathbf{r},\mathbf{r}^{\prime })>g^{(2)}(\mathbf{r},\mathbf{r})$.
Note that a non-ideal atom detection efficiency adds a
factor of $\eta<1$ to the second term in Eq.~(\ref{V-g}), but does not affect the
Cauchy-Schwartz inequality.

An immediate consequence of Eq.~(\ref{V-g}) is that the observed
number-difference squeezing can be insignificant if the pixel
occupation number $\hat{N}_{\mathbf{r}}$ is very small due to a
small pixel size, even if the pair
correlation $g^{(2)}(\mathbf{r},\mathbf{r}^{\prime })$ is much larger than $%
g^{(2)}(\mathbf{r},\mathbf{r})$. In other words, fine spatial
resolution can degrade the number-difference squeezing
between the elementary pixels, and larger pixel size or binning  favor
observing strong squeezing.

\begin{figure}[tbp]
~~\includegraphics[height=3.34cm]{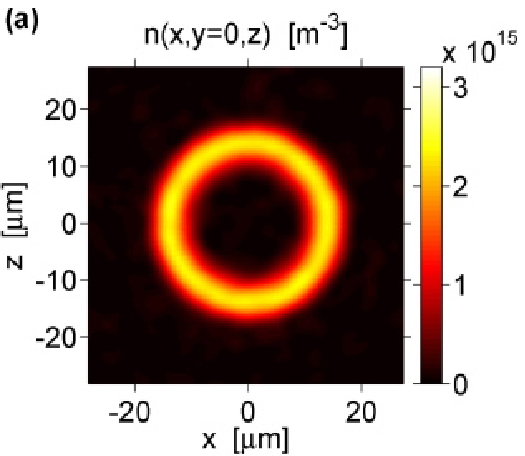}~\includegraphics[height=3.38cm]{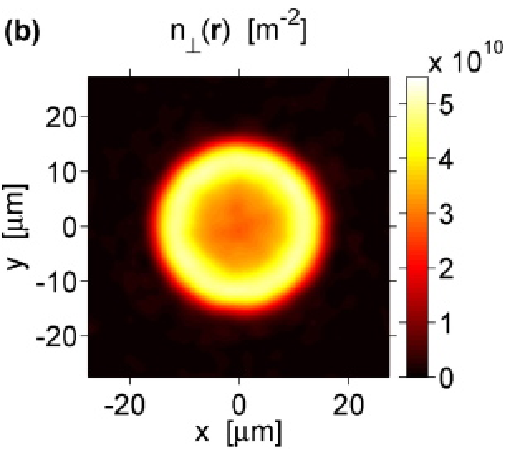}\\
\includegraphics[height=3.36cm]{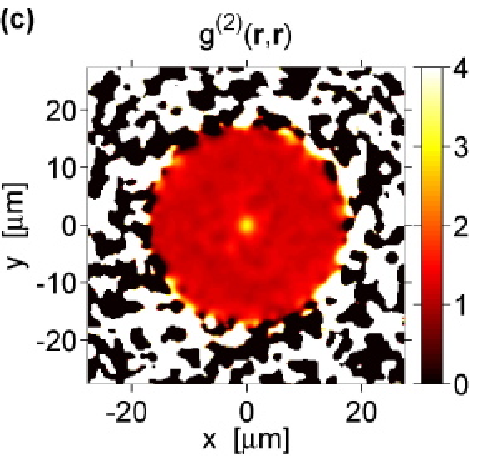}~~~\includegraphics[height=3.36cm]{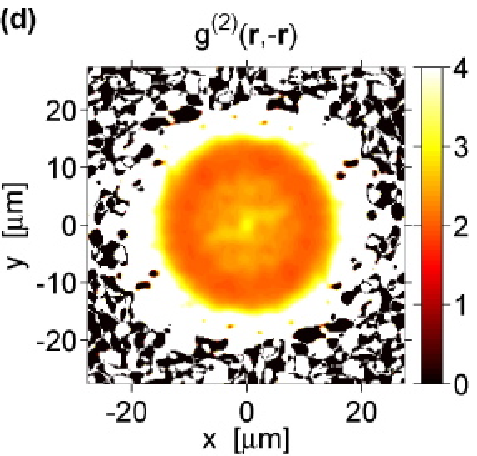}~~~\\
~\includegraphics[height=3.36cm]{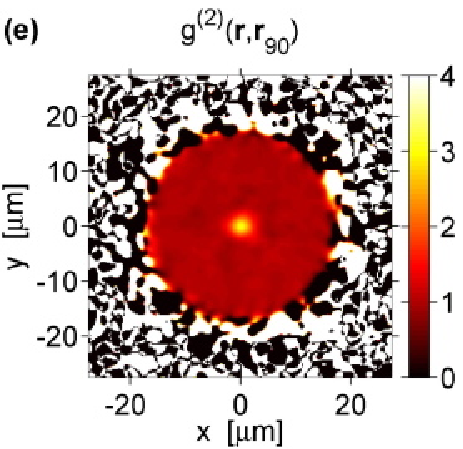}~~~~\includegraphics[height=3.34cm]{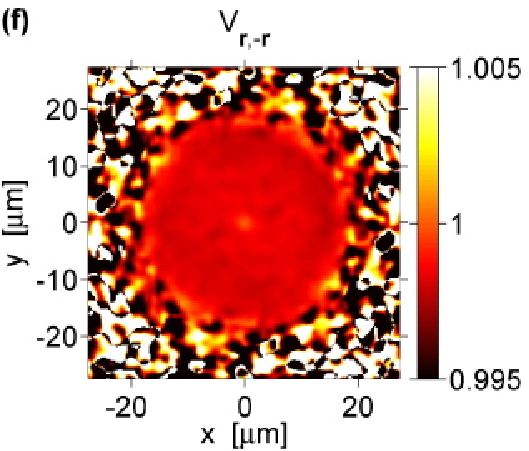}\\
\includegraphics[height=3.36cm]{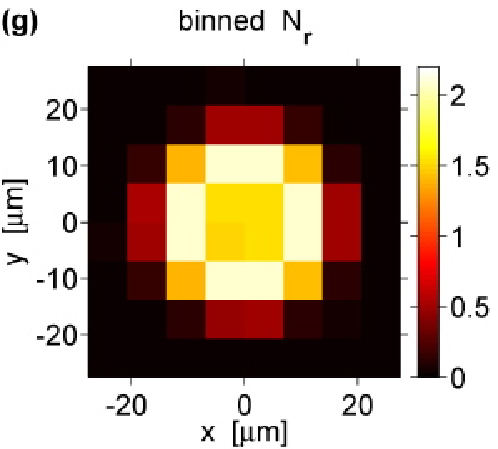}~\includegraphics[height=3.36cm]{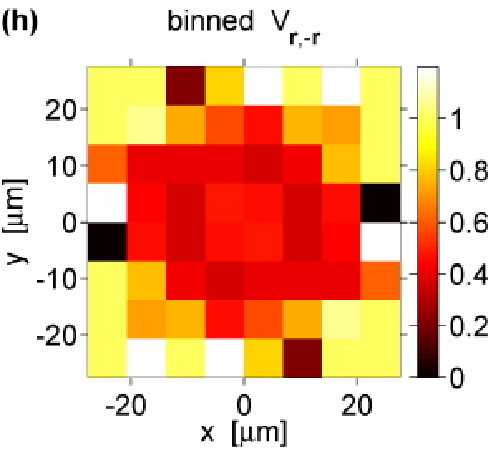}
\caption{(color online) Dissociation of a molecular
condensate
in free space, for dissociation time of $t=0.2$ ms with $\protect%
\chi =$ $7\times 10^{-7}$ m$^{3/2}$ s$^{-1}$ and a further expansion of $%
t_{e}=2.5$ ms with $\protect\chi =0$  \protect\cite{parameters}.
Each dimension is $55$ $\protect\mu$m long and
contains $256$ lattice points. $10,000$ stochastic paths were
averaged. (a) Slice through $z=0$ of the 3D
atomic density $n(\mathbf{x})=\langle \hat{n}(\mathbf{x})\rangle $
after expansion.
(b) 2D column density $n_{\perp }(\mathbf{r})=\langle \hat{n}_{\perp }(%
\mathbf{r})\rangle $ after integration along the $z$ axis. (c) local
pair-correlation, $g^{(2)}(\mathbf{r},\mathbf{r})$. The $g^{(2)}$ data is too
noisy outside the dissociation sphere where the signal is negligible
and should be discarded at radii $r\gtrsim 15$ $\protect\mu$m. (d)
Pair-correlation at
opposite locations, $g^{(2)}(\mathbf{r},-\mathbf{r})$. (e) Pair-correlation $%
g^{(2)}(\mathbf{r},\mathbf{r}_{90})$ at orthogonal locations. (f)
Number-difference variance $V_{\mathbf{r},-\mathbf{r}}$ at opposite
locations. (g) Binned atomic signal on the detection plane, for bins
of size $32\times 32$ pixels, and (h) number-difference variance
between the opposite bins.} \label{Figure1}
\end{figure}

Fig.~\ref{Figure1} shows the results of a simulation of the
dissociation
of a small molecular BEC of size $\sim 1.4$ $\mu$m, containing about $%
205$ molecules (in subsequent examples we simulate larger
condensates, containing $874$ and $9.2\times 10^{3}$ molecules
\cite{parameters}). The images relate to the atomic field after
dissociation and expansion. There are about $30$ atoms in total in
this example, corresponding to $7$\% conversion.
Fig.~\ref{Figure1}(a) shows a slice through the 3D atomic density,
which forms an expanding spherical shell about $14$ $\mu$m in
radius. Fig.~\ref{Figure1}(b) shows the 2D column density after
integration along the $z$ axis. The projection of the spherical
shell of atoms is clear and is similar to the experimental
observations of Refs.~\cite{Durr,Greiner}.
Figs.~\ref{Figure1}(c)-(d) show correlations in column densities at
the same, opposite, and orthogonal locations: (c) shows the local
pair correlation, $g^{(2)}(\mathbf{r},\mathbf{r})$,
which has a thermally bunched character in 3D, $g^{(2)}(\mathbf{x},\mathbf{x%
})=2$~\cite{moldisspaper1}, while the obtained value of $g^{(2)}(\mathbf{r},\mathbf{%
r})\simeq 1.2<2$ is due to the integration along $z$; (d) shows
strong pair-correlation at opposite locations,
$g^{(2)}(\mathbf{r},-\mathbf{r})\simeq 2.05 >
g^{(2)}(\mathbf{r},\mathbf{r})$, originating from the
momentum correlations of dissociated atom pairs; and (e) shows
an uncorrelated signal of $g^{(2)}(\mathbf{r},\mathbf{r}_{90})\simeq 1$, where $%
\mathbf{r}_{90}$ corresponds to a $90^{\circ}$ rotation of the
original image about the origin. In~(a)-(f), the elementary pixel
size is given by the computational grid, with $0.215$ $\mu$m spacing
in each dimension. The numerical convergence of our simulations is
ensured by smaller grid sizes reproducing the results within the
stochastic sampling errors.

In the example of Fig.~\ref{Figure1} we have a violation of the
classical
Cauchy-Schwartz inequality as $g^{(2)}(\mathbf{r},-\mathbf{r})>g^{(2)}(%
\mathbf{r},\mathbf{r}) $. However, due to small occupation numbers of the
elementary pixels, $\langle \hat{N}_{\mathbf{r}%
}\rangle $, the associated number-difference variance at the
diametrally opposite locations, $V_{\mathbf{r},-\mathbf{r}}$, shows
very little squeezing in Fig.~\ref{Figure1}(f),\ as explained by Eq.
(\ref{V-g}); it deviates from $V_{\mathbf{r},-%
\mathbf{r}}=1$ by only $-1.7\times 10^{-3}$ on the projected
dissociation sphere where the signal is maximal.
Figs.~\ref{Figure1}(g) and (h) show that measurement bins of size
$32\times 32$ pixels improve the number difference squeezing to
$V_{\mathbf{r},-\mathbf{r}}\simeq 0.4$ ($60$\% squeezing).

\begin{figure}[tbp]
\includegraphics[height=7cm]{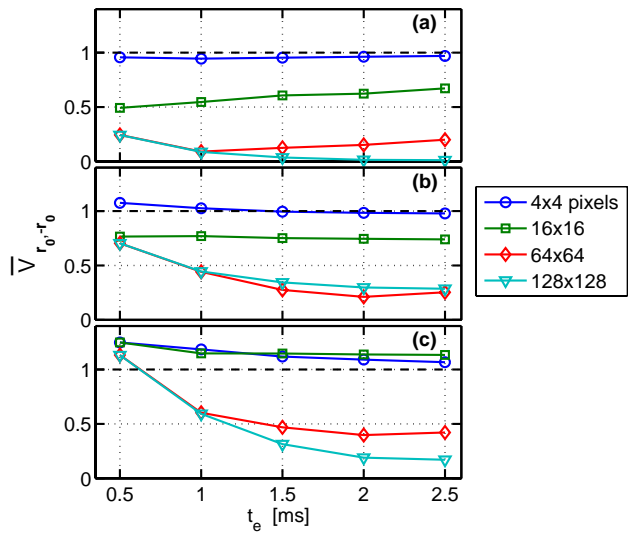}
\caption{(color online) The effect of binning on the angle averaged
number-difference variance between the opposite bins,
$\overline{V}_{\mathbf{r_{0}},-\mathbf{r_{0}}}$, as a function of
the expansion time $t_{e}$ following $0.2$ ms of dissociation.
Graphs (a), (b) and (c) correspond, respectively, to the initial
molecular BEC size of $1.4$, $3$ and $6.4$ $\mu $m
\cite{parameters}. The different curves are for different bin sizes.
The largest bins into just four quadrants on the detection plane.}
\label{Figure2}
\end{figure}

To understand the dependence of the correlation strength on the
initial size of the molecular BEC and the expansion time, we
analyzed the binned number-difference variance for three different
cases. Fig.~\ref{Figure2} shows the dependence of the angle averaged
variance $\overline{V}_{\mathbf{r_{0}},-\mathbf{r_{0}}}$ on the
expansion time, for different bin sizes. At each time, the angle
averaged result is at the radius $r_{0}$ of the dissociation sphere,
where the signal is maximal.

As we see from Fig.~\ref{Figure2}, the squeezing improves with
larger bin sizes in all cases, consistent with the
experimental results of Ref.~\cite{Greiner}. For the case of a small condensate,
Fig.~\ref{Figure2}(a), the squeezing degrades with the expansion time
due to the large momentum uncertainty of the initial molecular BEC.
The resulting center-of-mass momentum offset of
correlated atom pairs, causes them to fail to appear in diametrically
opposite bins in the far field. Increasing the bin size captures the
pairs in the opposite (larger) bins and restores the pair
correlations. For a larger molecular condensate, as in
Fig.~\ref{Figure2}(c), the complimentary effect is that the position
uncertainty may give the atom pairs a center-of-mass position
offset, again preventing them from appearing in the diametrically
opposite bins in the near field. In this case, the squeezing
improves with expansion, corresponding to a more complete conversion
of the intrinsic opposite-momentum correlations into spatial
correlations in the far field. The case of Fig. \ref{Figure2}(b) is
intermediate and is affected by the competition between these two
effects, resulting in the optimum expansion time and implying the
existence of an optimum size of the molecular BEC for a given
expansion time.

The authors acknowledge stimulating discussions with A. Perrin and
C. Westbrook, and thank the developers of the XMDS software (see
www.xmds.org). The work was supported by the Australian Research
Council.


\begin{thebibliography}{99}
\bibitem{parameters} Fig. \ref{Figure1} is for the initial
molecular BEC in a spherically symmetric harmonic trap of frequency $%
\omega /2\pi =90$ Hz, with a density profile given by the
ground-state of the Gross-Pitaevskii equation, obtained for
$^{87}$Rb$_{2}$ dimers at a scattering length of $2$ nm, as an
example. The chemical potential is chosen to give a peak density of
$5\times 10^{19}$ m$^{-3}$, which results in a total of $205$
molecules and a half-width of $\sim 1.4$ $\mu$m. In the two
additional examples of Fig.~\ref{Figure2}, the trap frequencies are
$\omega /2\pi =45$ Hz and $19$ Hz, while the chemical potentials are
adjusted to result in the same peak density, giving a total of $874$
and $9.2\times
10^{3}$ molecules, respectively, and half-widths of $\sim 3$ $\mu$m and $%
\sim 6.4$ $\mu$m. The detuning is $\Delta /2\pi =-3$ kHz in all
cases.
\end{thebibliography}
\end{document}